\documentclass[journal,10pt]{IEEEtran}
\usepackage{graphicx,psfrag,subfigure,url,amsmath,amsfonts,
epsfig,amssymb}

\newtheorem{conj}{Conjecture}

\newtheorem{theorem}{Theorem}

\newtheorem{lemma}{Lemma}
\newtheorem{remark}{Remark}
\newtheorem{bound}{Bound}

\def\cW{{\mathcal W}}
\newcommand{\p}{{\rm P}}
\long\def\symbolfootnote[#1]#2{\begingroup
\def\thefootnote{\fnsymbol{footnote}}\footnote[#1]{#2}\endgroup}

\begin{document}

\title{On the inner and outer bounds for 2-receiver discrete memoryless broadcast channels}
\author{Chandra Nair, CUHK and Vincent Wang Zizhou, CUHK}

\maketitle

\begin{abstract}
We study the best known general inner bound\cite{mar79} and outer bound\cite{nae07} for the capacity region of the two user discrete memory less channel. We prove that a seemingly stronger outer bound is identical to a weaker form of the outer bound that was also presented in \cite{nae07}. We are able to further express the best outer bound in a form that is computable, i.e. there are bounds on the cardinalities of the auxiliary random variables.

The inner and outer bounds coincide for all channels for which the capacity region is known and it is not known whether the regions described by these bounds are same or different. We present a channel, where assuming a certain conjecture backed by simulations and partial theoretical results, one can show that the bounds are different.
\end{abstract}

\section{Introduction}
\label{se:intro}
In \cite{cov72}, Cover introduced the notion of a broadcast channel through which one sender transmits information to two or more receivers. For the purpose of this paper we focus our attention on broadcast channels with precisely two receivers.

{\em Definition: } A {\em broadcast channel} (BC) consists of an
input alphabet $\mathcal{X}$ and output alphabets $\mathcal{Y}_1$
and $\mathcal{Y}_2$ and a probability transition function
$p(y_1,y_2|x)$. A $((2^{nR_1}, 2^{nR_2}),n)$ code for a broadcast
channel consists of an encoder
\[ x^n : 2^{nR_1} \times 2^{nR_2} \rightarrow \mathcal{X}^n, \]
and two decoders
\[ \hat{\cW}_1 : \mathcal{Y}_1^n \rightarrow 2^{nR_1} \]
\[ \hat{\cW}_2 : \mathcal{Y}_2^n \rightarrow 2^{nR_2}. \]

The probability of error $P_e^{(n)}$ is defined to be the
probability that the decoded message is not equal to the transmitted
message, i.e., 

\[ P_e^{(n)} = \mathbf{P}\left(\{\hat{\cW}_1(Y_1^n) \neq \cW_1 \}\cup \{
\hat{\cW}_2(Y_2^n) \neq \cW_2 \} \right)
\]

where the message is assumed to be uniformly distributed over
$2^{nR_1} \times 2^{nR_2}$. 

A rate pair $(R_1,R_2)$ is said to be {\em achievable } for the
broadcast channel if there exists a sequence of $((2^{nR_1},
2^{nR_2}),n)$ codes with $P_e^{(n)} \rightarrow 0$. The {\em
capacity region} of the broadcast channel with is the closure of the set of achievable rates.
{\em 
 The capacity region of the two user discrete memoryless channel is unknown.}

The capacity region is known for lots of special cases such as degraded, less noisy, more capable, deterministic, semi-deterministic, etc. - see  \cite{cov98} and the references therein.

General inner and outer bounds for the two-user discrete memoryless broadcast channel have also been known in literature. Here we state the best known inner and outer bounds for the region from the literature. 
\begin{bound}
 \label{th:mar79ib}
 [M\"{a}rton '79] The following rate pairs are achievable:
 \begin{align*}
  R_1 &\leq I(U,W;Y_1) \\
  R_2 &\leq I(V,W;Y_2) \\
  R_1 + R_2 &\leq \min\{I(W;Y_1),I(W;Y_2)\} + I(U;Y_1|W) \\
  &\qquad + I(V;Y_2|W) - I(U;V|W)
 \end{align*}
 for any $p(u,v,w,x)$ such that $(U,V,W) \to X \to (Y_1, Y_2)$ form a Markov chain.
\end{bound}
\begin{bound}
 \label{th:nae07ob}
 [Nair-El Gamal '07] The region $\mathcal{R}$ defined by the union over the rate pairs satisfying
 \begin{align*}
  R_1 &\leq I(U,W;Y_1) \\
  R_2 &\leq I(V,W;Y_2) \\
  R_1 + R_2 &\leq \min\{I(U,W;Y_1) + I(V;Y_2|U,W), \\
  & \qquad I(V,W;Y_2) + I(U;Y_1|V,W) \}
 \end{align*}
 over all $p(u)p(v)p(w,x|u,v)$ such that $(U,V,W) \to X \to (Y_1, Y_2)$ form a Markov chain forms an outer bound 
 to the capacity region.
\end{bound}

\begin{remark}
 Both the bounds are tight for all the special classes of two-user broadcast channels for which the capacity region is known. However, since the bounds are difficult to evaluate in general it is not known whether the tightness of these bounds is specific to the scenarios or whether they coincide yielding the capacity region.
\end{remark}

A {\em possibly} weaker form of the outer bound was also presented in \cite{nae07} by removing the independence between $U$ and $V$. Under this relaxation we have the following:
\begin{bound}
 \label{th:nae07ob1}
 [Nair-El Gamal '07]  The region $\mathcal{R}_1$ defined by the union over the rate pairs satisfying
 \begin{align*}
  R_1 &\leq I(U;Y_1) \\
  R_2 &\leq I(V;Y_2) \\
  R_1 + R_2 &\leq \min\{I(U;Y_1) + I(V;Y_2|U), \\
  & \qquad I(V;Y_2) + I(U;Y_1|V) \}
 \end{align*}
 over all $p(u,v,x)$ such that $(U,V) \to X \to (Y_1, Y_2)$ form a Markov chain constitutes an outer bound to the capacity region. 
\end{bound}

One of the main results of the paper is the following: {\em The regions described by Bounds \ref{th:nae07ob} and \ref{th:nae07ob1} are identical.}

The organization of the paper is as follows. In Section \ref{se:genrem} we show that the regions described by Bound \ref{th:nae07ob} and Bound \ref{th:nae07ob1} are the same. We also present a different representation of the the bound which allows us to have bounds on the cardinalities of the auxiliary random variables. In Section \ref{se:bssc} we study the binary skew-symmetric channel \cite{hap79} and conjecture that the inner and outer bounds are different for this channel.

\section{On evaluation of the outer bound}
\label{se:genrem}

\subsection{Identity of the bounds}
\label{sse:ob}

\begin{theorem}
\label{th:regcoin}
 The regions $\mathcal{R}$ and $\mathcal{R}_1$ coincide, i.e. $\mathcal{R}=\mathcal{R}_1$. 
\end{theorem}

\begin{proof}
 Clearly, by setting $U'=(U,W)$ and $V'=(V,W)$, we have that $\mathcal{R} \subseteq \mathcal{R}_1$. Therefore it suffices to show that $\mathcal{R}_1 \subseteq \mathcal{R}$.
 
The idea of the proof \footnote{The idea of the construction is motivated in part by a similar construction \eqref{eq:constr} originally appearing in \cite{nae07} and also from a conversation with Prof. Hajek about the tightness of Bound \ref{th:nae07ob} for the deterministic broadcast channel.} is as follows: Given a $(U,V)$ we will produce a $(U^*,V^*,W^*)$ with $U^*,V^*$ being independent such that
\begin{align}
I(U;Y_1)&=I(U^*,W^*;Y_1) \nonumber \\
I(V;Y_2)&=I(V^*,W^*;Y_2) \nonumber \\
I(U;Y_1|V)&=I(U^*;Y_1|V^*,W^*) \label{eq:const}\\
I(V;Y_2|U)&=I(V^*;Y_2|U^*,W^*). \nonumber
\end{align}

Let $(U,V,X)$ be a triple such that $(U,V) \to X \to (Y_1,Y_2)$ form a Markov chain. Let $\mathcal{V} = \{0,1,...,m-1\}$. Define new random variables $U^*, V^*, W^*$  and a distribution $p(u^*,v^*,w^*,x)$ according to
 \begin{align*}
  &\p(U^*=u, V^* = i, W^* = j, X=x) \\
   & \quad = \frac{1}{m} \p(U=u,V=(i+j)_m,X=x), 
 \end{align*}
where $(\cdot)_m$ denotes the $\mod$ operation.

It is straightforward to check the following: 
\begin{align*}
 \p(U^*=u,V^*=i) &=\frac{1}{m}\p(U=u) \\
 &\quad\mbox{and hence independent}, \\
 \p(U^*=u,W^*=i,X=x) &=  \frac{1}{m}\p(U=u, X=x), \\
 \p(V^*=i,W^*=j,X=x) &=  \frac{1}{m}\p(V=(i+j)_m, X=x).
\end{align*}

From the above it follows in a straightforward manner that \eqref{eq:const} holds and thus
completes the proof.
\end{proof}

\subsection{An alternate characterization}
\label{sse:altob}

We reproduce some of the arguments in \cite{nae07} to express the Bound  \ref{th:nae07ob1} in an alternate manner to aid its evaluation.

\begin{lemma}
 \label{le:eqbd}
 The region $\mathcal{R}_1$ is equivalent to the following region, $\mathcal{R}_2$, defined by the union of rate pairs satisfying
 \begin{align*}
  R_1 &\leq I(U;Y_1) \\
  R_2 &\leq I(V;Y_2) \\
  R_1 + R_2 &\leq \min\{I(U;Y_1) + I(X;Y_2|U), \\
  & \qquad I(V;Y_2) + I(X;Y_1|V) \}
 \end{align*}
 over all $p(u,v,x)$ such that $(U,V) \to X \to (Y_1, Y_2)$
\end{lemma}

\begin{proof}
Since $(U,V) \to X \to (Y_1, Y_2)$ form a Markov chain, we have $I(U;Y_1|V) \leq I(X;Y_1|V)$ and $I(V;Y_2|U) \leq I(X;Y_2|U)$. Therefore it is clear that $\mathcal{R}_1 \subseteq \mathcal{R}_2$. Hence it suffices to show that $\mathcal{R}_2 \subseteq \mathcal{R}_1$.

Let $l$ denote the size of $\mathcal{X}$. Given $(U,V,X)$ it was shown in \cite{nae07} that for the following triple $(U^*,V^*,X)$, having cardinalities $l\|\mathcal{U}\|,l\|\mathcal{V}\|,l$ respectively, defined according to
\begin{equation}
\label{eq:constr}
\begin{aligned}
&\p(U^*=u_i,V^*=v_j)\\
&\quad \quad = \frac{1}{l}\p(U=u,V=v,X=(i-j)_l), \\
&\p(X^*=k|U^*=u_i,V^*=v_j)\\
& \quad \quad = \left\{
\begin{array}{ll} 1 &\mbox{if } k=(i-j)_l\\[2pt] 0 &\mbox{otherwise,}
\end{array} \right.
\end{aligned}
\end{equation}
one obtains
\begin{align}
I(U;Y_1)&=I(U^*;Y_1)  \nonumber \\
I(V;Y_2)&=I(V^*;Y_2)  \label{eq:eqbd3}\\
I(X;Y_1|V)&=I(X;Y_1|V^*) =  I(U^*;Y_1|V^*) \nonumber\\
I(X;Y_2|U)&=I(X;Y_2|U^*) = I(V^*;Y_2|U^*).\nonumber
\end{align}
Thus $\mathcal{R}_2 \subseteq \mathcal{R}_1$. \end{proof}

\subsection{Cardinality bounds}
\label{sse:cardbd}

Using the strengthened Carath\'{e}odory theorem by Fenchel and Eggleston
\cite{czk78} it can be readily shown that for any choice of the auxiliary random
variable $U$, there exists  a random variable $U_1$ with cardinality bounded
by $\| \mathcal{X} \| + 1$  such that $I(U;Y_1) = I(U_1;Y_1)$,
$I(X;Y_2|U) = I(X;Y_2|U_1)$ and preserves the distribution $p(X)$. Similarly one can find a $V_1$ with 
cardinality bounded by $\| \mathcal{X} \| + 1$  such that $I(V;Y_2) = I(V_1;Y_2)$,
$I(X;Y_1|V) = I(X;Y_1|V_1)$ and preserves the distribution $p(X)$. Since both $U_1$ and $V_1$ share the same distribution $p(X)$ one can create a triple $(U_1,V_1,X)$ (for e.g. by generating $U_1$ and $V_1$ conditionally independent of $X$). Thus one can assume without loss of generality that the cardinalities of $U,V$ in Lemma \ref{le:eqbd} are bounded by $\|\mathcal{X}\|+1$ each. 

\subsection{An outer bound formulation that can be evaluated}
\label{sse:evoubd}

Putting all of these  together we have the following characterization of the Bound  \ref{th:nae07ob}.
\begin{bound}
 \label{le:eqbd4}
 The region $\mathcal{R}$ consists of the  union of rate pairs satisfying
 \begin{align*}
  R_1 &\leq I(U;Y_1) \\
  R_2 &\leq I(V;Y_2) \\
  R_1 + R_2 &\leq \min\{I(U;Y_1) + I(X;Y_2|U), \\
  & \qquad I(V;Y_2) + I(X;Y_1|V) \}
 \end{align*}
 over all $p(u,v,x)$ such that $(U,V) \to X \to (Y_1, Y_2)$ and constitutes the Bound  \ref{th:nae07ob}. Further, one can assume that $\| \mathcal{U} \|, \| \mathcal{V} \| \leq \| \mathcal{X} \| + 1 $. 
 
 Alternately one can also use construction \eqref{eq:constr} to restrict $X$ to be a deterministic function of $U,V$ while relaxing the cardinalities to $\| \mathcal{U} \|, \| \mathcal{V} \| \leq \| \mathcal{X} \|\left(\| \mathcal{X} \| + 1\right). $
\end{bound}
 
\section{The binary skew-symmetric channel}
\label{se:bssc}

\subsection{On evaluating M\"{a}rton inner bound}

We consider the following channel \cite{hap79} called the Binary skew-symmetric channel, BSSC. For ease we restrict ourselves to the case $p= \frac 12$.

\begin{figure}[ht]
\begin{center}
\begin{psfrags}
\psfrag{X}[r]{$X$}
\psfrag{Y_1}[l]{$Y_1$}
\psfrag{Y_2}[l]{$Y_2$}
\psfrag{p}[b]{$p$}
\psfrag{1-p}[c]{$1-p$}
\psfrag{0}[c]{$0$}
\psfrag{1}[c]{$1$}
\includegraphics[width=0.75\linewidth,angle=0]{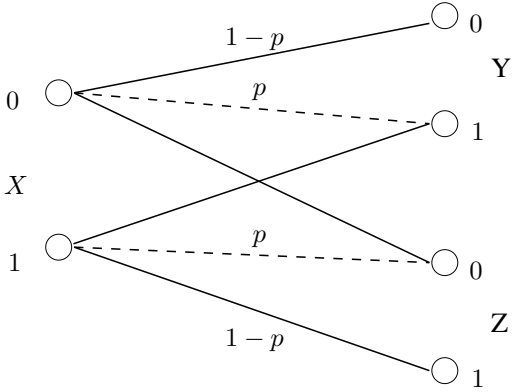}
\end{psfrags}
\caption{Binary Skew
Symmetric Channel}
\label{fig:bssc}
\end{center}
\end{figure}

\begin{remark}
The channel, BSSC, has already appeared in a couple of instances to produce
the following surprising results:
\begin{itemize}
\item In \cite{hap79} BSSC was used to show that using the auxiliary random variable $W$ in the Cover-van der Meulen achievable region, even in the absence of rate $R_0$ (common information), enhanced the achievable region. 
\item In \cite{nae07} BSSC was used to show that an outer bound to 2-user broadcast channel by Korner and M\"{a}rton \cite{mar79} was not tight and that the region prescribed by Theorem \ref{th:nae07ob} was strictly contained inside the Korner-M\"{a}rton region.
\end{itemize}
\end{remark}

\medskip

Backed by numerical simulations we make the following conjecture about the BSSC with $p=\frac 12$.
\begin{conj}
\label{co:sum}
Let $(U,V)$ be auxiliary random variables such that $(U,V) \to X \to (Y_1,Y_2)$ form a Markov chain. Then the following holds:
$$I(U;Y_1) + I(V;Y_2) - I(U;V) \leq \max\{ I(X;Y_1), I(X;Y_2) \}.$$ 
\end{conj}

\begin{remark}
\label{re:td}
It is easy to see that this conjecture implies that Marton's bound without the random variable $W$ reduces to the time-division region.
\end{remark}

When $U$ and $V$ are independent, this conjecture has been established in the appendix of \cite{hap79}. In this paper, we shall establish the validity of the conjecture for some ranges of $\p(X=0)$. 

By symmetry of BSSC the maximum of the term $I(U;Y_1) + I(V;Y_2) - I(U;V)$ is same for $\p(X=0)=\eta$ and $\p(X=0)=1-\eta$ and hence it suffices to consider $\eta$ in the range $0 \leq \eta \leq \frac 12$.

Observe that 
\begin{align*}
& I(U;Y_1) + I(V;Y_2) - I(U;V), \\
& \quad \leq I(V;Y_2) + I(U;Y_1,V) - I(U;V), \\
& \quad = I(V;Y_2) + I(U;Y_1|V), \\
& \quad \leq I(V;Y_2) + I(X;Y_1|V), \\
& \quad = I(X;Y_1) + I(V;Y_2) - I(V;Y_1).
\end{align*}

\begin{figure}[ht]
\begin{center}
\includegraphics[width=0.8\linewidth,angle=0]{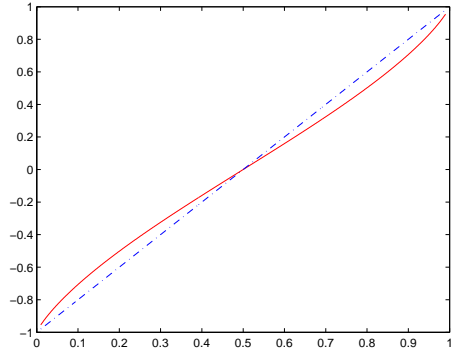}
\caption{The plot of the function $f(\eta) = H(\frac{\eta}{2}) - H(\frac{1-\eta}{2})$.}
\label{fig:diff}
\end{center}
\end{figure}

Figure \ref{fig:diff} plots $H(Y_1)-H(Y_2)$ and the line $2\eta-1$ as a function of $\p(X=0)=\eta$.
Let $f(\eta) = H(\frac{\eta}{2}) - H(\frac{1-\eta}{2})$, where $H(\cdot)$ denotes the binary entropy function. 
Then it is easy to see that $f(\eta)$ 
is concave in $0\leq \eta \leq \frac{1}{2}$ and convex in the remaining region, $\frac{1}{2} \leq \eta \leq 1$.

Suppose that $\p(X=0)=\eta$ and we seek the $V$ that maximizes $I(V;Y_2)-I(V;Y_1)$ subject to $V \to X \to (Y_1,Y_2)$ being Markov and $\p(X=0)=\eta$. Then it is not difficult to see that the optimal choice would be to set $V=\Phi$ (the trivial random variable) for all $\eta \leq \eta_0=  \frac 15$ where $\eta_0$ is the unique  point in $[0,\frac 12]$ at which the line joining $(\eta_0, f(\eta_0))$ to the point $(1,1)$ is a tangent to the curve $f(\eta)$.

\begin{lemma}
 \label{le:Vmax}
 Let $\p(X=0) = \eta \leq \eta_0 $ where $\eta_0= \frac 15$ is the unique solution of the equation
$$ f'(\eta) = \frac{1-f(\eta)}{1-\eta}. $$ 
or in other words the point at which the line joining $(\eta_0, f(\eta_0))$ to the point $(1,1)$ is a tangent to the curve $f(\eta)$. 

Then for all $V \to X \to (Y_1,Y_2)$ we have $I(V;Y_2) \leq I(V;Y_1)$.
\end{lemma}

\begin{proof}
 Define $g(\eta)$ as follows:
 $$ g(\eta) = \begin{cases} f(\eta) & 0 \leq \eta \leq \eta_0 \\ \frac{1-\eta}{1-\eta_0} f(\eta_0) + \frac{\eta-\eta_0}{1-\eta_0}f(1) & \eta_0 \leq \eta \leq 1 \end{cases}. $$
 Observe that $g(\eta)$ is concave and that $f(\eta) \leq g(\eta), 0 \leq \eta \leq 1$.

Let $\p(V=i) = v_i$ and $\p(X=0|V=i) = \alpha_i$. We have $\sum_i v_i \alpha_i = \eta$. Observe that we have the following,
\begin{align*}
 &I(V;Y_2) - I(V;Y_1) \\
 & \quad = H(Y_1|V) - H(Y_2|V) - \left( H(Y_1) - H(Y_2) \right) \\
 & \quad = \sum_i v_i f(\alpha_i) - f(\eta) \\
 & \quad \leq \sum_i v_i g(\alpha_i) - f(\eta) \\
 & \quad \stackrel{(a)}{\leq}  g(\sum_i v_i \alpha_i) - f(\eta) \\
 & \quad = g(\eta) - f(\eta) = 0 ~\mbox{as}~ 0 \leq \eta \leq \eta_0.
\end{align*}
Here $(a)$ follows from the concavity of $g(\eta)$. This completes the proof of Lemma \ref{le:Vmax}.
\end{proof}

This implies that for $\eta \leq \eta_0 = \frac 15$, we have
\begin{align*}
& I(U;Y_1) + I(V;Y_2) - I(U;V), \\
& \quad \leq I(X;Y_1) + I(V;Y_2) - I(V;Y_1), \\
& \quad \leq I(X;Y_1) + 0, \\
& \quad = I(X;Y_1).
\end{align*}

Further using the symmetry of BSSC and the fact that  the maximum of $I(U;Y_1) + I(V;Y_2) - I(U;V)$ is same for $\p(X=0) = \eta~\mbox{or}~1-\eta$, we have the following result.
\begin{lemma}
\label{le:cotrue}
Conjecture \ref{co:sum} is true as long as 
$\max\{\p(X=0),\p(X=1)\} \leq \eta_0 = \frac 15.$
\end{lemma}

Assuming Conjecture 1 is true we can now analyze the sum rate of the Marton inner bound with the random variable $W$. Theorem \ref{th:mar79ib} implies
\begin{align*}
R_1 + R_2 &\leq \min\{I(W;Y_1),I(W;Y_2)\}  \\
&\quad  + I(U;Y_1|W) + I(V;Y_2|W) - I(U;V|W). 
\end{align*}

Let $\mathcal{W}_0 = \{w: \p(X=0|W=w) \leq 0.5 \}$ and $\mathcal{W}_1 = \{w: \p(X=0|W=w) > 0.5 \}$. Let $T$ be a function of $W$ defined by
\begin{equation*}
T= \begin{cases} 0 &\mbox{if}~w\in \mathcal{W}_0 \\ 1 & \mbox{if}~w\in \mathcal{W}_1 \end{cases}.
\end{equation*}
We have the following bound on the sum rate
\begin{align*}
R_1 + R_2 &\leq \min\{I(W,T;Y_1),I(W,T;Y_2)\} \\
&\quad  + I(U;Y_1|W,T) + I(V;Y_2|W,T) \\
&\qquad  - I(U;V|W,T) \\
& \stackrel{(a)}{\leq} \min \{I(W,T;Y_1),I(W,T;Y_2)\} \\
&\quad  + \p(T=0)I(X;Y_1|W,T=0) \\
&\qquad+ \p(T=1)I(X;Y_2|W,T=1)\\
& \stackrel{(b)}{\leq} \min\{I(T;Y_1),I(T;Y_2)\} \\
&\quad + \p(T=0)I(X;Y_1|T=0) \\
&\qquad + \p(T=1)I(X;Y_2|T=1).
\end{align*} 
Here $(a)$ follows from Conjecture \ref{co:sum} and $(b)$ follows from the fact that
\begin{align*}
& \p(T=1)I(W;Y_1|T=1) \leq \p(T=1)I(W;Y_1|T=1), \\
& \p(T=0)I(W;Y_2|T=0) \leq \p(T=0)I(W;Y_1|T=0).
\end{align*}

In \cite{nae07} the bound on sum rate, $ \min\{I(T;Y_1),I(T;Y_2)\}$ + $\p(T=0)I(X;Y_1|T=0)$ + $\p(T=1)I(X;Y_2|T=1)$ has been studied  and the maximum was evaluated as $\approx 0.3616$. This could also be inferred from \cite{hap79} and the evaluation of the Cover-van-der-Meulen region for this channel. 

{\em Thus assuming Conjecture \ref{co:sum} we have that the sum rate of the M\"{a}rton inner bound is bounded by $0.3616...$ (correct to 4 decimal places).}

\subsection{Evaluating outer bound - BSSC}
In \cite{nae07}  the sum rate of the pairs $(R_1,R_2)$ described by Bound \ref{le:eqbd4} was evaluated and it was shown that the maximum sum rate was bounded by $0.3711..$ (correct to 4 decimal places). {\em Thus  we have that the region described by Bound  \ref{th:nae07ob} is strictly larger than that described by Bound \ref{th:mar79ib}(assuming Conjecture 1)  and thus the inner and outer bounds differ for BSSC.}

\section{Conclusion}
\label{se:concl}
In this paper, we study the inner and outer bounds for the 2-user discrete memoryless broadcast channel. We prove that for the purpose of evaluating the outer bound the region described by a {\em weaker} version (which is easier to evaluate) indeed coincides with a stronger version.

The bounds matched for all the special classes of channels for which the capacity was known. It is not known if the bounds were inherently different or not. We then studied the bounds for the particular case of the binary skew symmetric channel (BSSC).
We present a conjecture that, if proved, would establish that the inner and the outer bounds are indeed not tight for BSSC. Numerical simulations also indicate that the bounds differ for BSSC.

This definitely indicates that one of the bounds or possibly both are weak. We have demonstrated that resolving the capacity region for the BSSC would definitely give a strong hint on the capacity region of the broadcast channel for two users.

\section*{Acknowledgments}
The authors would like to acknowledge Prof. Bruce Hajek for very stimulating discussions during his visit to CUHK. The authors would also like to acknowledge some valuable suggestions and stimulating exchanges on the broadcast channel and on BSSC  by Prof. Abbas El Gamal.

\bibliographystyle{IEEEtran}
\bibliography{mybiblio}
\end{document}